\documentclass{webofc}

\usepackage[varg]{txfonts}   
\usepackage{hyperref}
\usepackage{graphicx} 
\usepackage{url}
\usepackage{lineno}
\hypersetup{colorlinks=true,citecolor=blue,urlcolor=blue,linkcolor=blue}
\begin{document}

%
\title{A Comprehensive Bandwidth Testing Framework for the LHCb Upgrade Trigger System}
%
%

\author{\firstname{Luke} \lastname{Grazette}\inst{1} \and
        \firstname{Ross} \lastname{Hunter}\inst{1}\fnsep\thanks{email: ross.john.hunterATcern.ch} \and
        \firstname{Ella} \lastname{Noomen}\inst{2} \and
        \firstname{Nicole} \lastname{Skidmore}\inst{1} \and
        \firstname{Sascha} \lastname{Stahl}\inst{3} \and
        \firstname{Mika} \lastname{Vesterinen}\inst{1} \and
        \firstname{Shunan} \lastname{Zhang}\inst{4}
}

\institute{
    Department of Physics, University of Warwick, Coventry, United Kingdom
    \and
    Faculty of Mathematics and Computer Science, Ruprecht-Karls-Universit{\"a}t Heidelberg, Heidelberg, Germany
    \and
    European Organization for Nuclear Research (CERN), Geneva, Switzerland
    \and
    Department of Physics, University of Oxford, Oxford, United Kingdom
}

\abstract{The LHCb experiment at CERN has undergone a comprehensive upgrade, including a complete re-design of the trigger system into a hybrid-architecture, software-only system that delivers ten times more interesting signals per unit time than its predecessor. This increased efficiency - as well as the growing diversity of signals physicists want to analyse - makes conforming to crucial operational targets on bandwidth and storage capacity ever more challenging. To address this, a comprehensive, automated testing framework has been developed that emulates the entire LHCb trigger and offline-processing software stack on simulated and real collision data. Scheduled both nightly and on-demand by software testers during development, these tests measure the online- and offline-processing's key operational performance metrics (such as rate and bandwidth), for each of the system's 4000 distinct physics selection algorithms, and their cumulative totals. The results are automatically delivered via concise summaries - to GitLab merge requests and instant messaging channels - that further link to an extensive dashboard of per-algorithm information. The dashboard and pages therein facilitate test-driven trigger development by 100s of physicists, whilst the concise summaries enable efficient, data-driven decision-making by management and software maintainers. This novel bandwidth-testing framework has been helping LHCb build an operationally-viable trigger and data-processing system whilst maintaining the efficiency to satisfy its physics goals.}

\maketitle

\section{Introduction}
\label{sec:intro}

The recent upgrade~\cite{UpgradeIPaper} of the Large Hadron Collider Beauty (LHCb) experiment included a full re-design of the trigger system: the first-level hardware trigger (L0) was removed, and thus LHCb became the first hadron-collider experiment to run with a full-software, heterogeneous-architecture trigger system processing events at 30 MHz. With more collisions per unit time in Run 3 (2022-2026) than Run 2 (2015-2018) and a more efficient trigger system, the expectation was that LHCb would be able to reconstruct $\sim10\times$ more signals of interest per unit time in Run 3 than in Run 2. 

\begin{figure}
    \centering
    \includegraphics[width=\linewidth]{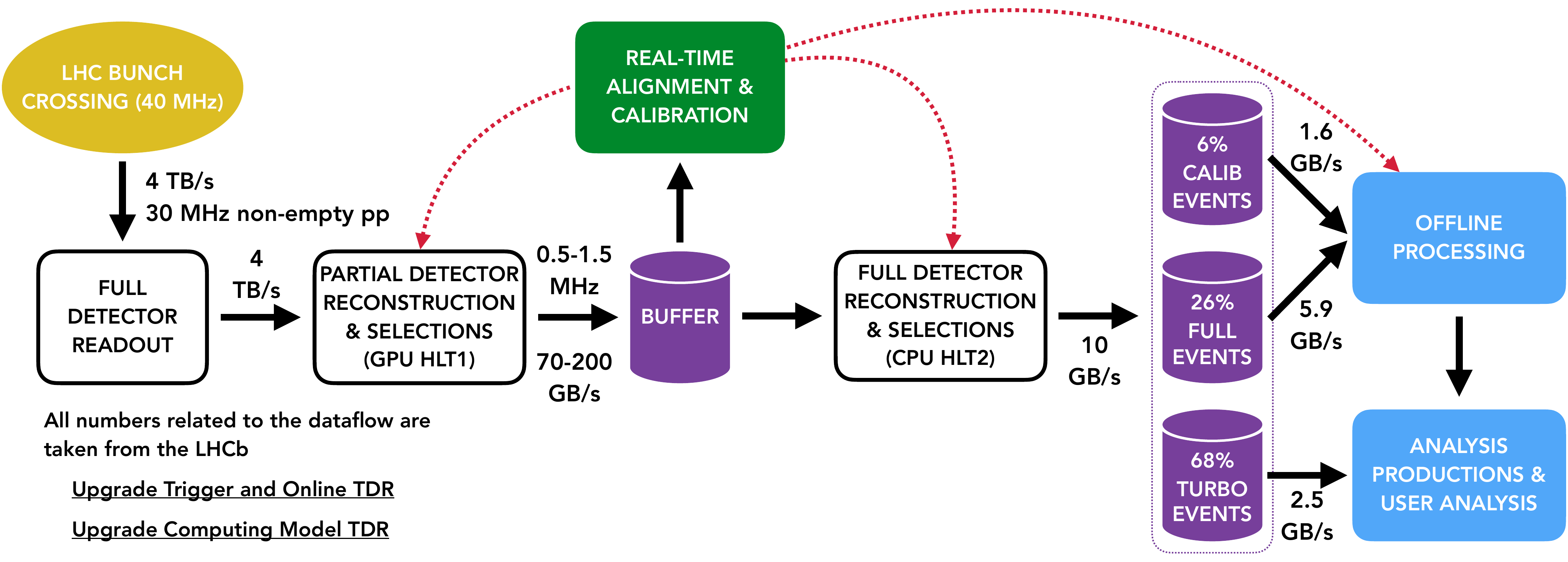}
    \caption{The data flow from collision events through the LHCb trigger and offline data-processing system. Note that ``offline processing'' here is also known as the ``Sprucing'' stage. Taken from Ref.~\cite{DataFlowFigure}.}
    \label{fig:dataflow}
\end{figure}

The flow of collision data through the trigger and data-processing system is shown in Fig.~\ref{fig:dataflow}. Visible collision events occurring at $\sim30$ MHz are reconstructed and filtered by a three-stage process. HLT1~\cite{Allen} takes raw detector data and performs a fast GPU-based partial reconstruction and filtering, which cuts the rate of events by $\sim 30\times$, outputting only interesting events which are fed (via a disk buffer) to HLT2. With the benefit of a lower input rate of events and a lower throughput\footnote{Events processed per second, typically measured in kHz.} requirement, HLT2 then performs a state-of-the-art reconstruction before filtering by more complex physics decay signatures. Output of HLT2 is then saved to permanent magnetic tape storage. Depending on the output \textit{stream} of HLT2, an event either undergoes minimal further processing (\texttt{Turbo} stream), or it passes through a final filtering stage known as ``Sprucing'' (\texttt{Full} stream)\footnote{Other streams also exist but are not the main focus here.}. The Sprucing acts as a safety net: the HLT2 \texttt{Full} stream filtering is typically looser than \texttt{Turbo}, and the tighter Sprucing can be periodically re-run if the final trigger strategy needs to be adjusted. However, this safety comes at an extra operational cost (to be explained in the next subsection), meaning that saving to \texttt{Turbo} is the default. Outputs of the offline processing are saved to permanent disk storage for physics analysis.

HLT1's main constraint is being able to process events quickly enough and to reduce the event \textit{rate} (events per unit time) to a manageable input for HLT2. HLT2 and the Sprucing are mainly constrained by \textit{bandwidth} (data saved to storage per unit time), since this quantity is proportional to the cost of purchasing and maintaining the required storage capacity. Bandwidth targets to tape and disk were set for Run 3 in the LHCb Upgrade Computing Model Technical Design Report (TDR)~\cite{UpgradeCompModel}, some of which are shown in Fig.~\ref{fig:dataflow}. These TDR targets are subject to change, but are indicative of the rough scale of the constraint. 

Unfortunately, monitoring how fast our storage fills up does not give sufficient information to meet this bandwidth constraint and maintain an efficient system. To help elucidate this, we define the bandwidth formally as

\begin{equation}\label{eq:bw_basic}
    \textrm{Bandwidth [MB/s]} = \textrm{Output Rate [kHz]} ~\times \textrm{Event Size [kB]} \, ,
\end{equation}

where the typical units have been given. An ideal trigger system would save only the interesting events (``signal'') and throw away the rest (``background''). A real trigger system cannot do this perfectly, and typically a looser trigger is more efficient at selecting signal but also lets in more background, and thus has a higher output rate. Trade-offs between signal efficiency and rate must therefore be made. In LHCb, the size of the events written out is also adjustable by persisting more or less information about the event~\cite{Tesla}. The minimal (default) option is to save only the trigger ``candidate'', e.g. two tracks from the same origin that are consistent with the decay of a $b$-hadron. The opposite extreme is to save all reconstructed quantities (tracks, calorimeter clusters etc.) in the event and/or the full raw readout of LHCb's detector. Events output to the HLT2 \texttt{Full} stream are required to save all reconstructed quantities so that the Sprucing can construct new trigger candidates when running over it. This explains why the Sprucing workflow cannot be our default in Run 3; its extra persistency results in extra bandwidth.

A further complication is that our system is not a single algorithm. Each trigger stage (HLT1, HLT2 and Sprucing) is a collection of trigger \textit{lines}. Each line is a selection algorithm itself, and is typically written and managed by a handful of analysts, has $\mathcal{O}(10)$ tunable selection parameters/thresholds, and has configurable persistency settings. For an event to pass each stage, at least one line must have been satisfied. HLT1 has $\mathcal{O}(10)$s trigger lines, and the available output rate is divided optimally and equitably between them via the minimization of a loss function~\cite{evans2025automatedbandwidthdivisionlhcb}. However, since there are more than 4,000 HLT2 and Sprucing lines - and thus tens of thousands of tunable parameters - this automated procedure cannot be repeated for the other stages. Instead, the LHCb management must liaise with line authors and set small-group targets to equitably divide the available bandwidth. Line authors must then tune their lines manually to hit these targets whilst maintaining sufficient signal efficiency.


To meet the bandwidth challenge with a system so complex requires substantial coordinated effort by a large number of people. Managers require top-level, cumulative information, whilst line authors need detailed information on their selection line to enable them to balance efficiency against bandwidth. Both groups need to know how these quantities change in response to software modifications. The aim of this work was to provide a comprehensive bandwidth-testing framework to meet these requirements and help LHCb build an efficient trigger and offline-processing system that meets its bandwidth targets.
\section{Implementation}
\label{sec:implementation}

\subsection{Workflow of a test}

\begin{figure}
    \centering
    \includegraphics[width=\linewidth]{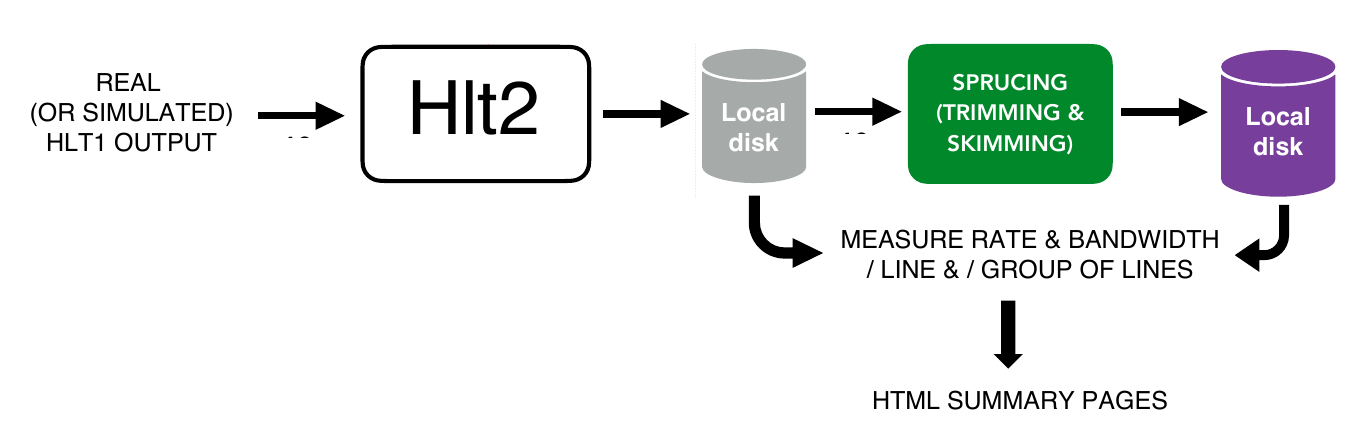}
    \caption{Schematic of the bandwidth-testing framework for LHCb's trigger and offline-processing system. The HLT2 and Sprucing applications are configured identically to real data-taking.}
    \label{fig:bwtest_workflow}
\end{figure}

The bandwidth-testing framework (just ``bandwidth tests'' in the following) is designed to emulate the real trigger and offline data processing workflow as closely as possible, but be able to run on a single machine with quick time-to-insight. The data flow of our main bandwidth test\footnote{For reference of LHCb readers, this is named \texttt{Moore\_hlt2\_and\_spruce\_bandwidth}.} is shown in Fig.~\ref{fig:bwtest_workflow}, to be compared with Fig.~\ref{fig:dataflow}. The bandwidth test starts from a fixed output of HLT1 (from input of either real or simulated collision events), which are fed into the HLT2 application configured identically to real data-taking. The HLT2 output is processed by the similarly-aligned Sprucing application, and both write to the local disk. Additionally, various metadata files are saved that document the trigger configuration and the input sample. With these metadata files and the HLT2 and Sprucing output files we have full flexibility to measure rate and bandwidth of each trigger stage in total, for each line, and any further groupings of lines. We can also measure related quantities, such as the overlap fraction or Jaccard similarity between pairs of lines/groups of lines.

The bandwidth of a stream is then (a re-expression of Eqn.~\ref{eq:bw_basic}):

\begin{equation}\label{eq:bw}
    \textrm{Stream Bandwidth} = \frac{\textrm{Input Rate} \times \textrm{Output File Size}}{N_{\rm input}} \, ,
\end{equation}

where $N_{\rm input}$ is the number of events the test ran over, and the input rate is equal to the output rate from the previous trigger stage (documented with the input file). The per-stream rate can be calculated accordingly as 

\begin{equation}\label{eq:stream_rate}
    \textrm{Stream Rate} = \textrm{Input Rate} \times \frac{N_{\rm output}}{N_{\rm input}} \, .
\end{equation}

To evaluate quantities per-line, specialised Python scripts iterate through the streamed output files event-by-event and access the ``decision reports'' which document the trigger lines that fired on that event. The per-line rate can then be calculated using Eqn.~\ref{eq:stream_rate} counting only events that fired that line. This is an \textit{inclusive} rate, in that it is inclusive of whether any other lines also fired on each event. We also calculate an \textit{exclusive} rate, which counts events where only the line in question fired towards its numerator $N_{\rm output}$. Comparing the two rates gives insight into the overlap of a line with other lines, which may be indicative of selections being too loose. The bandwidth attributed to each line is calculated by measuring the size of the reconstructed objects and raw-data objects present in those events in which the line in question fired. This is summed for all events in which the line fired and used in Eqn.~\ref{eq:bw} to calculate a per-line bandwidth. Note that this is an \textit{inclusive} bandwidth, and it should be interpreted carefully if multiple lines fire on the same events. In this case, the per-line bandwidth is misleadingly-high because candidates from each line will contribute to the event size. Two bandwidths are always measured: a total bandwidth and a ``DstData'' bandwidth, where only the container of reconstructed information is counted towards the event size in the latter. Comparing the two bandwidths gives insight into the impact of a line's persistency settings. 

Measured quantities are written out per-stream and per-line to \texttt{csv} intermediary files. With the Pandas data analysis library~\cite{pandas1,pandas2} these raw tables are then formatted, sorted and/or re-grouped for presentation. For ease of interpretation a series of plots are also made, such as bar charts of per-stream bandwidths and pie charts of lines grouped by similar physics signatures. The final step of the workflow is to collate these plots and tables into a single HTML document to be added to a web-based dashboard.

The most important variant of the bandwidth tests has been discussed here, but there are currently six bandwidth tests that test different trigger stages with different input files. The workflow is broadly the same in each case.

\subsection{Scheduling of the tests}

The bandwidth tests are scheduled to run on a dedicated test machine by the LHCb Performance and Regression~\cite{LHCbPR} interface to the Jenkins automation software~\cite{Jenkins}. Each test nominally runs every night from a nightly build of the LHCb trigger and offline-processing software stack on the default \texttt{git} branches. The software stack is hosted in a series of GitLab~\cite{gitlab} projects, and so before a merge request (e.g. adding new lines or modifying the reconstruction) is accepted, a continuous integration (CI) test is ordered by the software shifters/maintainers. This test builds the relevant part of the stack with the code changes included before running a suite of fast integration and regression tests. Addition of an appropriate label to a merge request will signal Jenkins (via a GitLab web-hook) to schedule a bandwidth test on the merge-request build as part of the CI test.  

Jenkins does not run the workflow outlined in the previous subsection directly, but rather calls a ``handler'' Python script. The handler runs the bandwidth test as a sub-process, handles logging and tracks performance statistics such as sub-job exit codes, time taken and memory usage. The handler also searches the logs for hints of problems. All relevant plots, HTML code and intermediary files are copied to an accessible space in which the HTML code will be deployed to a web dashboard. The final job of the handler is to post succinct summary messages (via API) to a Mattermost~\cite{mattermost} instant-messaging channel, giving the most important bandwidth-related quantities, a success/failure message, and a link to the dashboard (and logs) for full information. One message is generated per trigger sub-job (roughly split by streams or stages). This message is carefully formulated for quick parsing e.g. by shifters or management. In the case of a bandwidth test triggered from a merge request, summary messages also contain a one-line comparison of the bandwidth changes, and they are also posted (via an API) to the GitLab merge request webpage. The comparison is measured with respect to the latest nightly bandwidth test, and comprehensive tables of how each metric has changed for each line and stream are also added to the dashboard. The hyperlinks from the automated messages open up the HTML dashboard in a browser. A chronological view (descending in time from the most recent to the oldest) of all bandwidth tests - with links to each - can also be seen from a top-level website. 

Each bandwidth test takes around 1 hour to report results from 100,000 input events on a 16-core test machine of modest specification. More input events would mean more precise estimates but a slower test. With an input rate of around 1 MHz (nominal HLT1 output rate), 1/100,000 events passing the trigger stages is 10 Hz, which is therefore our minimum output rate sensitivity. This is precise enough for the majority of lines, and ample precision for the cumulative per-stream and per-stage totals. 
\section{Example results and discussion}
\label{sec:discussion}

\begin{figure}[h]
    \centering
    \includegraphics[width=0.9\linewidth]{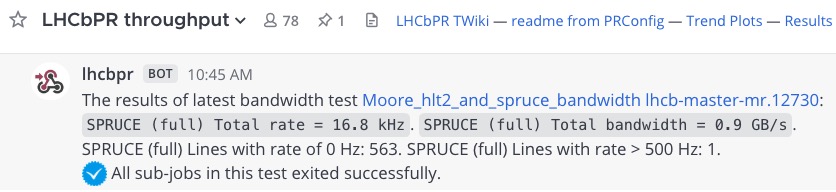}
    \caption{An example screenshot of automated feedback messages to a Mattermost instant-messaging channel. In this case the test was successful. The most important quantities are summarised, and a link to the dashboard is given.}
    \label{fig:mmost_feedback}
\end{figure}
\begin{figure}
    \centering
    \includegraphics[width=\linewidth]{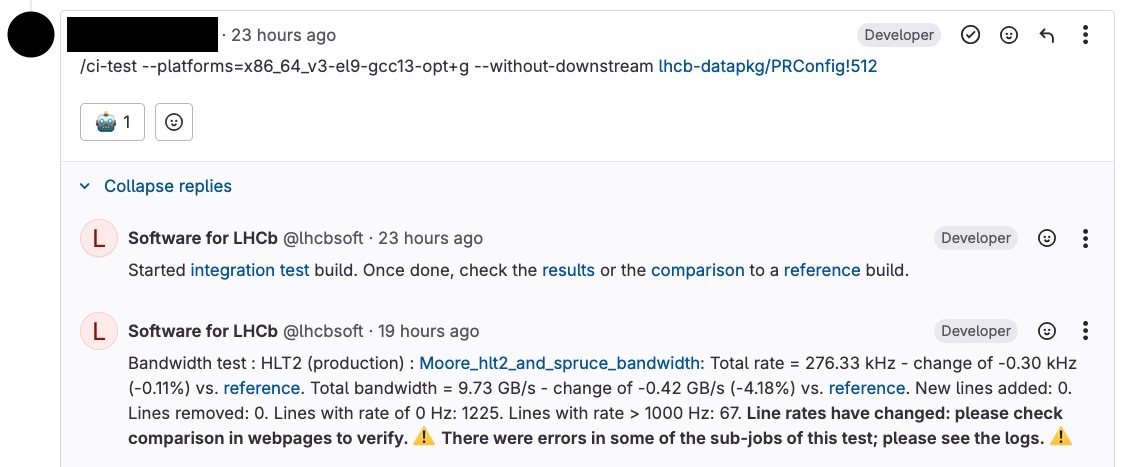}
    \caption{An example screenshot of automated bandwidth-test feedback messages on a GitLab merge request discussion. The name and picture of the user triggering the test has been redacted. In this example, the test had logged some errors and line rates changed with respect to the reference, so readers are prompted to follow-up.}
    \label{fig:gitlab_feedback}
\end{figure}
\begin{figure}
    \centering
    \includegraphics[width=0.95\linewidth]{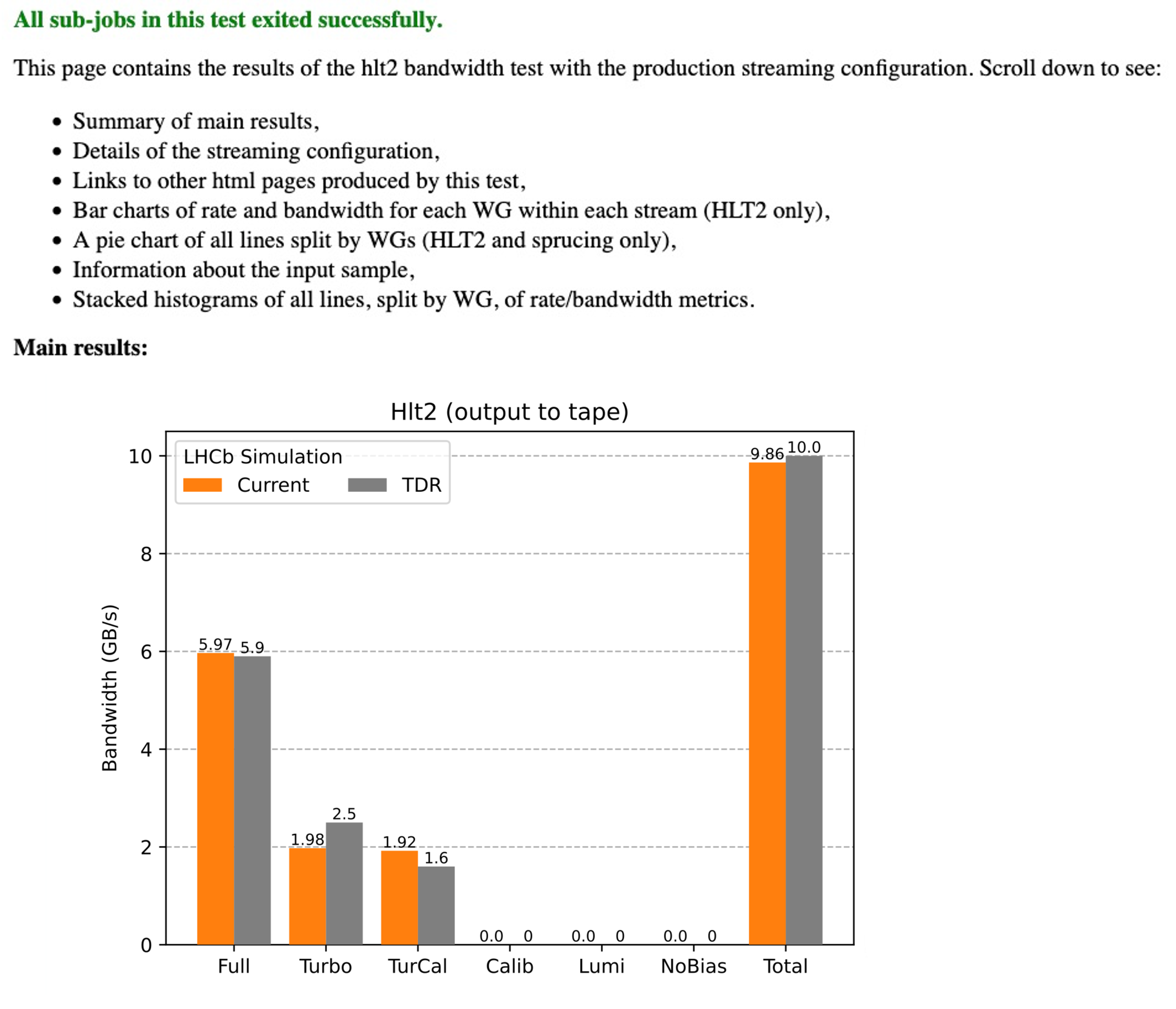}
    \caption{Example screenshot of the front-page of the bandwidth test webpages, or ``dashboard''. The most important metrics measured by the test are shown in the bar chart in orange, and compared to a target (grey) taken from Ref.~\cite{UpgradeCompModel}. The status of the test is noted, as well as what further results the reader can expect to find in the pages. In this instance simulated collision data was used as an input to the test. These results are also published as an example in Ref.~\cite{LHCB-FIGURE-2024-034}.}
    \label{fig:dashboard}
\end{figure}

Examples of the automated feedback messages posted to the Mattermost instant-messaging channel and a GitLab merge request are shown in Figs.~\ref{fig:mmost_feedback} and ~\ref{fig:gitlab_feedback}. The examples are picked to show cases of a successful and an unsuccessful test. This instant feedback at completion means that no monitoring of the test is necessary. Failure messages indicate either problems in the bandwidth tests themselves, the underlying software stack, or (most commonly) the changes being made in the merge request. The one-line comparison helps us make well-informed choices on whether a merge request can be accepted in terms of bandwidth. An example front-page of the dashboard is shown in Fig.~\ref{fig:dashboard}. Care has been taken here to keep the pages simple and to show the most important information first, before directing users to progressively more in-depth information as they move through the web pages linked therein. The final layer of approval on a GitLab merge request is often based on scrutinizing the bandwidth test webpages, in particular their comparison pages.

\begin{figure}[ht]
\centering
\includegraphics[width=0.75\linewidth]{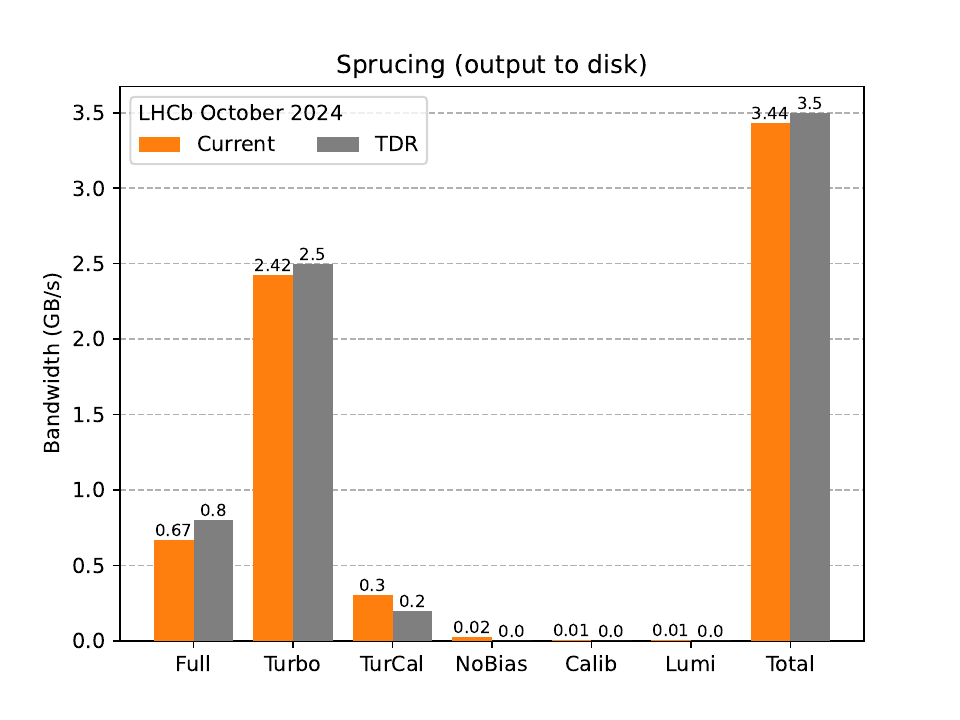}
\caption{Bandwidth of all streams to disk measured on real data taken in October 2024 (orange), as compared to the targets set in the LHCb Upgrade Computing Model Technical Design Report (TDR) (grey). The cumulative bandwidths of all streams is shown in the ``Total'' bar. Taken from Ref.~\cite{LHCB-FIGURE-2024-034}.}
\label{fig:bw_to_disk}
\end{figure}

Since their deployment in 2023, the bandwidth tests have been used routinely to test changes to our software stack. The nightly management-level view has motivated action when necessary to fit within bandwidth constraints. The culmination of this consistent bandwidth-driven decision-making can be seen in Fig.~\ref{fig:bw_to_disk}, which shows a snapshot of the bandwidth to disk as measured on real collision data (at our nominal Run 3 instantaneous luminosity) in October 2024. The measured bandwidth is compared to the targets mentioned earlier from Ref.~\cite{UpgradeCompModel}, and can be seen to satisfy them. This was all the more impressive in that HLT1 had been deliberately loosened to allow 20\% more output. The substantial achievement is not only that we can hit these targets, but that the system also has a high efficiency of selecting events for physics analysis. The full proof of this achievement will be in upcoming physics measurements from LHCb, although it already visible in other contributions from LHCb in these proceedings. This bandwidth-testing framework has played a major role in facilitating these achievements. 
 
Looking forward, work is ongoing to integrate plots of the kinematic distributions (e.g. transverse momentum of a combination of tracks) of candidates that each line is selecting. This promises to drastically speed up selection refinement and the response time if bandwidths need to be adjusted.
\section{Conclusion}
\label{sec:conclusion}

Keeping the new LHCb trigger and offline-processing system within operational limits on bandwidth, whilst also being efficient enough to satisfy physics goals, presents a considerable challenge. Measuring enough bandwidth-related information to overcome this challenge necessitates a comprehensive bandwidth-testing framework. The tests described here emulate the full trigger and offline-processing system on a single test machine, and output in the form of concise instant-messaging summaries with links to extensive web-based dashboards. The hierarchical presentation of results means that everyone from the management to the author of each algorithm is well-informed on bandwidth. As demonstrated here, this has facilitated the construction of an operationally-viable trigger and offline-processing system. The tests continue to benchmark the impact of software improvements, and incoming additions to the tests promise to make the process of bandwidth-driven optimisation even faster still in preparation for 2025 LHCb data-taking.

\bibliography{main.bib} 

\end{document}